\documentclass[default,iicol]{sn-jnl}

\usepackage{graphicx}%
\usepackage{multirow}%
\usepackage{amsmath,amssymb,amsfonts}%
\usepackage{amsthm}%
\usepackage{mathrsfs}%
\usepackage[title]{appendix}%
\usepackage{xcolor}%
\usepackage{textcomp}%
\usepackage{manyfoot}%
\usepackage{booktabs}%
\usepackage{algorithm}%
\usepackage{algorithmicx}%
\usepackage{algpseudocode}%
\usepackage{listings}%
\usepackage{natbib}

\usepackage{breakurl}

\usepackage{scalerel}
\usepackage{tikz}
\usetikzlibrary{svg.path}

\definecolor{orcidlogocol}{HTML}{A6CE39}
\tikzset{
  orcidlogo/.pic={
    \fill[orcidlogocol] svg{M256,128c0,70.7-57.3,128-128,128C57.3,256,0,198.7,0,128C0,57.3,57.3,0,128,0C198.7,0,256,57.3,256,128z};
    \fill[white] svg{M86.3,186.2H70.9V79.1h15.4v48.4V186.2z}
                 svg{M108.9,79.1h41.6c39.6,0,57,28.3,57,53.6c0,27.5-21.5,53.6-56.8,53.6h-41.8V79.1z M124.3,172.4h24.5c34.9,0,42.9-26.5,42.9-39.7c0-21.5-13.7-39.7-43.7-39.7h-23.7V172.4z}
                 svg{M88.7,56.8c0,5.5-4.5,10.1-10.1,10.1c-5.6,0-10.1-4.6-10.1-10.1c0-5.6,4.5-10.1,10.1-10.1C84.2,46.7,88.7,51.3,88.7,56.8z};
  }
}

\newcommand\orcidicon[1]{\href{https://orcid.org/#1}{\mbox{\scalerel*{
\begin{tikzpicture}[yscale=-1,transform shape]
\pic{orcidlogo};
\end{tikzpicture}
}{|}}}}

\usetikzlibrary{backgrounds}
\makeatletter

\tikzset{%
  fancy quotes/.style={
    text width=\fq@width pt,
    align=justify,
    inner sep=1em,
    anchor=north west,
    minimum width=\linewidth,
  },
  fancy quotes width/.initial={.8\linewidth},
  fancy quotes marks/.style={
    scale=8,
    text=white,
    inner sep=0pt,
  },
  fancy quotes opening/.style={
    fancy quotes marks,
  },
  fancy quotes closing/.style={
    fancy quotes marks,
  },
  fancy quotes background/.style={
    show background rectangle,
    inner frame xsep=0pt,
    background rectangle/.style={
      fill=gray!20,
      rounded corners,
    },
  }
}

\newenvironment{fancyquotes}[1][]{%
\noindent
\tikzpicture[fancy quotes background]
\node[fancy quotes opening,anchor=north west] (fq@ul) at (0,0) {``};
\tikz@scan@one@point\pgfutil@firstofone(fq@ul.east)
\pgfmathsetmacro{\fq@width}{\linewidth - 2*\pgf@x}
\node[fancy quotes,#1] (fq@txt) at (fq@ul.north west) \bgroup}
{\egroup;
\node[overlay,fancy quotes closing,anchor=east] at (fq@txt.south east) {''};
\endtikzpicture}

\makeatother

\usepackage{makecell}

\newcolumntype{x}[1]{>{\centering\arraybackslash}p{#1}}

\makeatletter
\newcommand{\mycite}[2][]{%
  \textsuperscript{[\@bibitemShorthand{} as cited in \@bibitemLabel{}#1]}}
\makeatother

\theoremstyle{thmstyleone}%
\theoremstyle{thmstyletwo}%

\theoremstyle{thmstylethree}%

\begin{document}

\title[Exploiting the Margin: How Capitalism Fuels AI at the Expense of Minoritized Groups]{Exploiting the Margin: How Capitalism Fuels AI at the Expense of Minoritized Groups}


\author{\fnm{Nelson} \sur{Col\'on Vargas} \orcidicon{0009-0009-9038-7328}}



\affil{\orgdiv{Leverhulme CFI}, \orgname{University of Cambridge}, \orgaddress{\country{UK}}} 


\abstract{This paper explores the intricate relationship between capitalism, racial injustice, and artificial intelligence (AI), arguing that AI acts as a contemporary vehicle for age-old forms of exploitation. By linking historical patterns of racial and economic oppression with current AI practices, this study illustrates how modern technology perpetuates and deepens societal inequalities. It specifically examines how AI is implicated in the exploitation of marginalized communities through underpaid labor in the gig economy, the perpetuation of biases in algorithmic decision-making, and the reinforcement of systemic barriers that prevent these groups from benefiting equitably from technological advances. Furthermore, the paper discusses the role of AI in extending and intensifying the social, economic, and psychological burdens faced by these communities, highlighting the problematic use of AI in surveillance, law enforcement, and mental health contexts. The analysis concludes with a call for transformative changes in how AI is developed and deployed. Advocating for a reevaluation of the values driving AI innovation, the paper promotes an approach that integrates social justice and equity into the core of technological design and policy. This shift is crucial for ensuring that AI serves as a tool for societal improvement, fostering empowerment and healing rather than deepening existing divides.}




\maketitle

\pagebreak

\section*{Introduction}\label{sec1}
\begin{fancyquotes}
I know how to build a business. You gotta' get the black people to do it in order to get the white people to do it. Then you gotta' get the black people to stop doing it. --Dwight Shrute (The Office)
\end{fancyquotes}

\emph{The Office}, an American television series that aired from 2005 to 2013, is a critical mockumentary that satirizes the everyday lives of office employees. Achieving a peak viewership averaging over seven million viewers per episode during its most popular seasons and extensive syndication both domestically and internationally, the show became a ratings success and a cultural phenomenon. Its satirical approach not only influenced American perceptions of work culture but also reflected broader societal and cultural issues. The show's depiction of corporate management styles and mundane office life, as exemplified by the quote from character Dwight Shrute, serves as an entry point for discussing systemic racial and economic exploitations. This dynamic has long seen marginalized\footnote{``to relegate to an unimportant or powerless position within society or group'' \cite{pratt2022marginalized}.} groups as economic levers, initially utilized for their labor and subsequently sidelined to maintain a status quo that privileges a select few. The United States' economic ascent, fueled by the toil of enslaved individuals and later the exploitation of minority workers post-abolition, epitomizes this grim reality. Abraham Lincoln's admission in a letter to Horace Greeley \cite{lincoln1862}|prioritizing union preservation over emancipation|underscores a consistent preference for economic unity over racial justice.

\begin{fancyquotes}
If I could save the Union without freeing any slave, I would do it...[I]f I could save it by freeing some and leaving others alone, I would also do that. -Abraham Lincoln \cite{lincoln1862}
\end{fancyquotes}

This paper delves into the interplay between capitalism and racial discrimination, tracing its historical roots and its perpetuation through contemporary technologies like artificial intelligence (AI). Starting with Cedric Robinson's foundational ideas on racial capitalism, we explore how economic systems historically exploited racial distinctions to enhance capitalist gains, a pattern that is starkly mirrored in today's AI technologies. These systems, designed and deployed within the same capitalist framework, inherently manifest and amplify the biases and divisions that have been programmed into them, whether intentionally or inadvertently.

As we move into a detailed analysis of AI's role in modern society, the focus shifts to how this technology, while heralded as a tool of efficiency and progress, actually reproduces and exacerbates inequalities. This is evident in the labor practices within the tech industry, where AI development often relies on underpaid and undervalued workers from marginalized communities, perpetuating a cycle of exploitation and exclusion. Moreover, the deployment of AI in various sectors|from law enforcement to social services|raises significant concerns about fairness, transparency, and the potential for these systems to entrench social divides deeper. The discussion extends to the personal and societal impacts of these technologies, highlighting the real and often overlooked consequences of biased algorithms and system designs that prioritize profit over people.

Ultimately, this exploration is not just an academic exercise but a call to action. It underscores the urgent need for a critical reevaluation of how AI technologies are developed and employed. By integrating principles of social justice, equity, and inclusivity into the heart of AI development, we can begin to dismantle the structures of oppression that have long been sustained by capitalist endeavors. We advocate for transformative approaches that ensure AI serves as a force for good, promoting societal healing and empowerment rather than exacerbating the injustices of the past.

\section*{Historical Background}

The concept of racial capitalism, as analyzed by Cedric Robinson [\hyperref[robinson1983black]{Ro83} as cited in \hyperref[jenkins2021histories]{JL21}, p.5], provides a foundational framework for understanding how capitalism has historically operated by accentuating and institutionalizing pre-existing social distinctions, transforming them into racial differences. Jenkins and Leroy further explain, ``Racial capitalism is the process by which the key dynamics of capitalism|accumulation/dispossession, credit/debt, production/surplus, capitalist/worker, developed/underdeveloped, contract/coercion, and others|become articulated through race\footnote{Throughout this paper, when discussing a topic through the lens of race, we are working from Bonilla-Silva's definition of racism and systemic racism: ``Racism is the product of racial domination projects (e.g., colonialism, slavery, labor migration). Once these racial projects emerged in human history, racism became embedded in societies, that is, it became systemic racism.'' \cite[p.~20]{bonilla2003racism}.}.'' This mechanism of racial capitalism, while analyzed in the context of the United States, is not unique to it nor solely a dynamic between white and black populations. Globally, racial capitalism manifests wherever economic systems exploit racial distinctions for economic gain, regardless of the majority or minority status of the racial groups involved \cite{jenkins2021histories}. 

In the United States, the economic growth was notably founded on the exploitation of minoritized\footnote{``Minoritized populations are groups that are marginalized or persecuted because of systemic oppression.'' \cite{nih2024minoritized}} labor|particularly that of Black people and Indigenous peoples|a pattern that persisted from slavery through sharecropping, extracting profit by appropriating minority ingenuity while eroding their dignity. K-Sue Park's \cite{park2021race} account illustrates how Indigenous peoples lost land through strategies like foreclosure, capitalizing on their unfamiliarity with colonizers' economic practices. This pattern of exploitation and dispossession is a hallmark of racial capitalism, seen both in historical and contemporary settings.

Post-abolition, racial capitalism continued to dehumanize and underpay non-white workers, leading to persistent racial wealth gaps and exclusion. Pedro A. Regalado's \cite{regalado2021latinx} examination of Latinx entrepreneurship in the 1960s shows how economic opportunities inadvertently intensified internal divisions within the community, reflecting Robinson's insights into how capitalism amplifies pre-existing social distinctions such as legal status. Additionally, Regalado demonstrates how linguistic vulnerabilities were exploited, a tactic reminiscent of those employed against Indigenous communities centuries earlier, as discussed by Park.

In parallel, artificial intelligence emerges as a contemporary manifestation of these age-old disparities, capable of mirroring and magnifying biases embedded within the data it processes. Just as capitalism leveraged pre-existing social hierarchies for gain, AI risks doing the same, reinforcing racial disparities under the guise of technological advancement. In a sense AI technology, just as capitalism, functions as a machine that exploits existing patterns|in data|to accelerate production|of outputs. Scholars like Ruha Benjamin and Simone Browne caution against an unchecked AI, which could not only perpetuate but also amplify systemic oppressions through sophisticated means of profiling and control. This scenario underscores the imperative to critically examine AI through the prism of racial capitalism, recognizing the potential for these technologies to further entrench inequalities\footnote{The uneven and unfair distribution of opportunities and rewards that increase power, prestige, and wealth for individuals or groups; social disparity. \cite{dictionary}} rather than dismantle them.

\section*{AI: The Modern Frontier of Exploitation}\label{sec2}

AI doesn't operate in a vacuum; it mirrors the society that gives it life. The tech sector manufactures tools embedded with social biases, training AI models that `think' like a corporation|which are predominantly steered by white men \cite[p.35]{benjamin2019race}|at the expense of ambitions and morals \cite{penn2018ai}. This predilection for efficiency over human values carries profound implications for the implementation of AI-driven social services and their consequences for marginalized communities. The rapid proliferation of automated systems in these services raises concerns surrounding fairness, transparency, and accountability, driven by the relentless pursuit of efficiency. In this section, we aim to illustrate how capitalism's influence on AI systems consistently drives them to perpetuate disparities, whether in terms of labor exploitation, unequal time burdens, or the erosion of minority well-being, underscoring how marginalized communities are disproportionately affected across various dimensions of AI development.

\subsection*{Labor Tax: How Underpaying Minority Workers Fuels AI Development}\label{subsec1}

In their investigative piece \emph{The Exploited Labor Behind Artificial Intelligence}, Williams, Miceli, and Gebru \cite{williams2022exploited} expose the harsh realities faced by gig workers in the AI industry, who are disproportionately drawn from vulnerable populations. Big tech players in the AI industry such as Amazon and Facebook heavily rely on these laborers, including data labelers, content moderators, and delivery drivers. These workers, despite their indispensable contributions, often endure abysmal pay, constant surveillance, unrealistic quotas, and even safety hazards. Further details from the investigation reveal that data labelers, responsible for critical AI training data, can earn as little as \$1.77 per task. In the troubling case of content moderators in Kenya, they are relentlessly monitored to make split-second decisions, even when confronted with disturbing content. These workers, whose labor sustains billion-dollar companies, are unjustly compensated.

Izaguirre \cite{izaguirre2023uber} discussed a very recent example of labor exploitation within the gig economy from the ridesharing giants, Uber and Lyft. In New York, these companies are set to pay a staggering \$328 million to settle complaints that they wrongly imposed taxes and fees on their drivers that should have been covered by passengers. State Attorney General Letitia James unveiled this settlement, emphasizing how Uber and Lyft|which incorporate AI technologies for routing and pricing algorithms|had systematically deprived their drivers of millions of dollars in earnings and essential benefits. The drivers, akin to those engaged in data labeling for AI models, toiled for long hours in challenging conditions while being undercompensated. The repercussions of such labor exploitation have wide-ranging implications, demonstrating the persistence of these issues across different sectors of the gig economy. These drivers, often working long hours in challenging conditions, waited an arduous eight years for justice, emphasizing the enduring time burden in their battle for rights--we will explore the time dimension in the next subsection.

Moreover, racial capitalism also rears its head in other ways in generative media AI applications. Startups have built systems to synthesize fake profile photos by exploiting images of minority models without consent or payment. The marginalized provide raw material|their unvalued data and likenesses|to advance technologies from which they do not benefit. An example of this is the Argentina-based design firm, Icons8, which built generative AI systems to synthesize fake profile photos by exploiting minority likenesses \cite{harwell2020dating}. Though the models' images provided core training data, they received no payment for this labor, nor any share of the systems' profits derived from their inputs. Once again, capitalism fueled AI development by commodifying bodies. The models' data and likenesses became raw materials for tech advancement from which they did not equally gain. This case provides further evidence that we must scrutinize and address the human exploitation underlying much AI progress.

\subsection*{Time Surcharge: The Burden on Minorities to Fix Biased Systems}\label{subsec2}

In 2019, a pivotal federal study by NIST unveiled the entrenched racial biases within facial recognition algorithms, revealing higher false positive rates for Asian, African American, and Native American faces compared to white counterparts \cite{harwell2019federal}. This disparity is not merely a technical oversight but a reflection of a broader societal issue where fairness is often assessed through a predominantly white lens, perpetuating and amplifying harmful racial stereotypes. These algorithms, failing more frequently with Black and Asian individuals, not only reinforce the misconception of homogeneity within these communities but also echo Cedric Robinson's argument that capitalism morphs pre-existing social differences into racial disparities, using technology to further these divides.

These biases extend beyond race to gender and age, with algorithms showing elevated error rates for women, the elderly, and the young compared to middle-aged white men. Simone Browne, in {\it Dark Matters} \cite[p.~113]{browne2015dark}, elucidates how these systems perpetuate stereotypes, such as misclassifying Asian males as females and Black women as men, illustrating how AI models, like capitalism, disseminates racial and gender stereotypes on a massive scale under the guise of technological neutrality.

The case of Brandon Mayfield, as discussed by Browne, exemplifies how biases in AI systems can have real-world consequences. A veteran and a lawyer who had recently converted to Islam, Mayfield was erroneously matched by FBI algorithms to a fingerprint found on a bag containing a detonating device linked to the 2004 Madrid bombing \cite[p.~115]{browne2015dark}. Despite being one of twenty matches, his recent conversion to Islam and status as a veteran, both associated with prejudicial stereotypes, led authorities to single him out. Mayfield's personal life and beliefs were publicly exposed under a false accusation, and he remained entangled in legal battles to clear his name until November 2010. This wrongful identification underscores how minoritized groups not only bear the brunt of these biases but also face a ``Privacy Levy,'' paying a significant personal and financial price due to prejudiced technological systems.

This systemic bias embedded within AI technologies serves capitalist ends in several ways. Firstly, it perpetuates a cycle where the burden of correcting and navigating biased systems falls disproportionately on marginalized communities, consuming their time and resources|a ``Time Surcharge'' that benefits the status quo by maintaining existing power dynamics. Secondly, the deployment of such technologies, especially in law enforcement and surveillance, opens lucrative markets for tech companies, capitalizing on the state's desire to monitor and control, often at the expense of those already marginalized. In essence, these biased algorithms are not just products of existing societal inequalities; they are tools that capitalism exploits to further entrench its dominance, ensuring that the costs of innovation are borne by those least able to bear them, while the profits accrue to those already in positions of power.

It is important to point out that the concept of a ``Time Surcharge'' extends beyond the realm of AI and technological bias, reflecting a broader systemic issue that minoritized communities face across various bureaucratic and social systems. This enduring burden of time, as analyzed in the context of racialized administrative burdens by Ray, Herd, and Moynihan in \emph{Racialized Burdens: Applying Racialized Organization Theory to the Administrative State} \cite{ray2022racializedburden}, is often utilized as a means to punish and control, reinforcing poor living conditions. Similar to the ways in which biased algorithms require marginalized groups to navigate and correct flawed systems, these bureaucratic processes consume excessive amounts of time, serving as barriers that maintain social and economic inequalities. These practices are reminiscent of historical strategies designed to disenfranchise and exploit marginalized populations, perpetuating a cycle where the costs of bureaucratic inefficiency and technological innovation are disproportionately borne by those least equipped to handle them, while the systems in place continue to benefit those in power. 

The development and implementation of biased AI systems are stark manifestations of racial capitalism, where technological advancements are leveraged to amplify and profit from racial and social disparities. The need for a concerted effort to address these biases is not just a matter of technical accuracy but of challenging the capitalist structures that incentivize and benefit from the perpetuation of inequality.

\subsection*{The Wellbeing Toll: How AI Progress Erodes Minority Health}\label{subsec3}

The advancement of AI technology, hailed as a cornerstone of modern innovation, often conceals a human toll disproportionately borne by marginalized communities. D'Ignazio and Klein's observation \cite[p.~183]{dignazio2020data} that the tech industry relies on the precarious labor of older women of color, contrasts sharply with the privileged demographics of Silicon Valley. This labor force, essential to refining AI technologies, faces not just economic exploitation but significant mental health challenges, as Billy Perrigo's investigations \cite{perrigo2022inside, perrigo2023exclusive} into the conditions of data labelers reveal. Workers engaged in making AI systems safer endure traumatic content exposure for meager wages, suffering profound mental health impacts without adequate support. This reality underscores a capitalist calculus where AI progress is prioritized over the well-being of those who enable it, revealing a stark exploitation that benefits technological advancement and profit margins at a significant human cost.


In her work {\it Automating Inequality} \cite[p.~173]{eubanks2018automating}, Virginia Eubanks provides comprehensive evidence of the discriminatory outcomes resulting from automated social services. One striking example is the case of Angel and Patrick's family. Angel seeking help from a counselor and taking medication to manage her PTSD is undoubtedly a responsible decision aimed at becoming a better-equipped caregiver for her children. However, her use of public assistance|due to their socioeconomic status|to access these essential services has an unintended negative consequence. It affects the family's AFST (Allegheny Family Screening Tool) score|produced by a predictive model|, putting them at risk of having social services intervene and potentially remove their children. This situation raises a critical question: Why should it be a choice between the parents' well-being and the children's welfare? Instead, it should be an `and' rather than an `or,' emphasizing the importance of supporting families as a whole. Moreover, this scenario exemplifies the ``Privacy Levy" where the act of seeking private counseling and prescriptions wouldn't flag the same concerns, highlighting the disproportionate burden placed on marginalized communities.

These instances reflect a broader trend of ``algorithmic exploitation,'' where the relentless pursuit of efficiency and profit in AI development and application exacerbates social disparities. This exploitation is twofold: firstly, it leverages the labor of marginalized workers under deplorable conditions, enhancing AI capabilities while neglecting the workers' health and economic stability. Secondly, it deploys these technologies in ways that disproportionately harm minority communities, whether through surveillance, law enforcement, or social services, thereby reinforcing systemic inequalities.

As this section concludes, it's clear that the capitalist benefit derived from these practices is multifaceted, encompassing direct economic gains from reduced labor costs and expanded market dominance, as well as indirect advantages through the perpetuation of a socio-economic order that maintains a readily exploitable workforce. The prioritization of technological advancement and profit over human dignity and equity reflects a fundamental misalignment of values, where the potential of AI to serve the common good is compromised by a capitalist ethos that values profit above people. Addressing these issues demands a reevaluation of the objectives and ethics guiding AI development and deployment. It necessitates a shift towards a more equitable approach that recognizes the intrinsic value of all labor, respects the dignity of every individual, and prioritizes the well-being of marginalized communities. Only through such a transformative shift can we harness the potential of AI to contribute positively to society, rather than perpetuating the injustices of a capitalist system that exploits the vulnerabilities of the least among us.

\section*{Discussion and Conclusion}\label{sec13}

This comprehensive exploration reveals compelling evidence that contemporary AI development mirrors and perpetuates the exploitation and marginalization of minoritized groups witnessed under historical racial capitalism. AI not only reflects but also concentrates deep societal biases ingrained in our structures. Despite promises of progress, this emerging technology entrenches social divides, much like capitalism historically intensified differences into racialized hierarchies. Uncontrolled AI poses a significant threat, exacerbating current oppressions and expanding racial disparities across dimensions such as labor, rights, and well-being. Undervalued minority workforces fuel systems they cannot equally access or shape, while flawed data and algorithms covertly deny opportunities and resources. The relentless pursuit of AI innovation disregards the resulting trauma among marginalized builders.

Race and economic exploitation are always already deeply embedded within the framework of AI, inherently so because they are foundational to the capitalist structures within which AI is developed and deployed. So, what can we do? As Pratyusha Kalluri advises, we need to shift our inquiry from simply questioning whether AI is `good' or `fair' to probing how it shifts power. We must advocate for greater inclusion in shaping AI technologies, particularly involving those who have been historically excluded. By involving these communities directly in the creation and regulatory processes, we can begin to address the imbalances of power and ensure that AI development not only avoids perpetuating inequalities but actively contributes to social justice \cite{kalluri2020power}.

Confronting and reforming the underlying capitalist structures and ideologies that shape technological development is essential. This approach demands more than technological innovation|it requires a transformation of the values and priorities that guide this innovation. Moving beyond techno-solutionism, we must foster an AI development ethos that is deeply informed by social justice, historical consciousness, and a commitment to dismantling rather than perpetuating the systems of oppression embedded within our society.

The recent incident involving Google's Gemini chatbot illustrates the pitfalls of superficially and belatedly addressing racism and bias in AI development. Despite its intention to create diverse and inclusive imagery, Gemini produced outputs that were historically inaccurate and racially insensitive, masquerading as authentic historical representations \cite{grant2024gemini}. Often, by the time corrective measures are implemented, the foundational data and structures are already steeped in the biases they aim to eradicate. This situation is akin to attempting to cleanse water once it has already been poisoned at the source. Thus, addressing racism and bias in AI requires more than just post-hoc adjustments; it demands a fundamental overhaul of the underlying frameworks and systems.

In light of such incidents, it becomes imperative to champion AI development processes that are not only technologically sound but are also socially just and inclusive. Several entities are at the forefront of these efforts, contributing through research and advocacy to mitigate biases within AI systems. For example, the Algorithmic Justice League (\url{https://www.ajl.org/}) focuses on public engagement and the mitigation of bias in AI applications, highlighting the necessity of community involvement in technological oversight. Similarly, Data \& Society (\url{https://datasociety.net/}) conducts in-depth studies on the social implications of data-centric technologies, emphasizing the need for interdisciplinary approaches to understand and address the complexities of AI deployment. Additionally, the AI Now Institute (\url{https://ainowinstitute.org}) offers critical insights into the rights and liberties of populations affected by automation and AI, advocating for enhanced regulatory frameworks that ensure these technologies serve the public good.

Ultimately, the path forward must involve reimagining the role of technology in society. By fostering an AI development culture that values each individual's dignity and rights, we can harness the potential of these powerful tools to create a more just and equitable world. This transformative vision requires a collective commitment to dismantling the oppressive structures that underpin current technological practices, paving the way for innovations that empower and heal rather than divide and exploit. To truly mitigate the risk of perpetuating racial injustices through AI, confronting and reforming the underlying capitalist structures and ideologies that shape technological development is essential. This approach demands more than technological innovation|it requires a transformation of the values and priorities that guide this innovation. Moving beyond techno-solutionism, we must foster an AI development ethos that is deeply informed by social justice, historical consciousness, and a commitment to dismantling rather than perpetuating the systems of oppression embedded within our society.




\begin{thebibliography}{0000}

\bibitem[Be19]{benjamin2019race}
Benjamin, R \emph{Race After Technology}. Polity Press, 2019.


\bibitem[Bo22]{bonilla2003racism}
Bonilla-Silva, E. \emph{Racism Without Racists}. Sixth Edition, Rowman \& Littlefield, 2022.

\bibitem[Br15]{browne2015dark}
Browne, S. \emph{Dark Matters: On the Surveillance of Blackness}. Duke University Press, 2015.

\bibitem[DK20]{dignazio2020data}
D'Ignazio, C., \& Klein, L. (2020). \emph{Data Feminism}. The MIT Press.

\bibitem[E19]{eubanks2018automating}
Eubanks, V. \emph{Automating Inequality}. St. Martin's Press, 2018.

\bibitem[G24]{grant2024gemini}
Grant, N. Google Chatbot's A.I. Images Put People of Color in Nazi-Era Uniforms. \emph{The New York Times}, February 26, 2024\url{https://www.nytimes.com/2024/02/22/technology/google-gemini-german-uniforms.html}

\bibitem[H19]{harwell2019federal}
Harwell, D. Federal study confirms racial bias of many facial-recognition systems, casts doubt on their expanding use. \emph{The Washington Post}, December 19, 2019. \url{https://www.washingtonpost.com/technology/2019/12/19/federal-study-confirms-racial-bias-many-facial-recognition-systems-casts-doubt-their-expanding-use/}.

\bibitem[H20]{harwell2020dating}
Harwell, D. Dating apps need women. Advertisers need diversity. AI companies offer a solution: Fake people. \emph{The Washington Post}, January 7, 2020. \url{https://www.washingtonpost.com/technology/2020/01/07/dating-apps-need-women-advertisers-need-diversity-ai-companies-offer-solution-fake-people/}.

\bibitem[I23]{izaguirre2023uber}
Izaguirre, A. (2023, November 2). Uber and Lyft to pay \$328 million to settle dispute over taxes and fees paid by New York drivers. \emph{AP News}. \url{https://apnews.com/article/uber-lyft-new-york-city-wage-theft-9ae3f629cf32d3f2fb6c39b8ffcc6cc6}.

\bibitem[JL21]{jenkins2021histories}
\label{jenkins2021histories}
Jenkins, D., \& Leroy, J. (Eds.). \emph{Histories of Racial Capitalism}. Columbia University Press, 2021.

\bibitem[K20]{kalluri2020power}
Kalluri, P. Don't Ask if AI is Good or Fair, Ask how it Shifts Power. \emph{Nature} Vol. 583, 2020.

\bibitem[Li62]{lincoln1862}
Lincoln, A. \emph{Letter in Reply to Horace Greeley on Slavery and the Union}. The Restoration of the Union the Paramount Object, August 22, 1862. Retrieved from \url{https://www.presidency.ucsb.edu/documents/letter-reply-horace-greeley-slavery-and-the-union-the-restoration-the-union-the-paramount}.

\bibitem[NIH]{nih2024minoritized}
Race and National Origin. In \emph{NIH Style Guide}. National Institutes of Health. Accessed April 24, 2024 \url{https://www.nih.gov/nih-style-guide/race-national-origin}.


\bibitem[O24]{dictionary}
\emph{Open Education Sociology Dictionary.} Accessed April 24, 2024. \url{https://sociologydictionary.org/inequality/#types_of_inequality}.

\bibitem[Pa21]{park2021race}
Park, K-Sue. ``1. Race, Innovation, and Financial Growth: The Example of Foreclosure". In \emph{Histories of Racial Capitalism}, edited by Justin Leroy and Destin Jenkins, 27-52. New York Chichester, West Sussex: Columbia University Press, 2021. \url{https://doi.org/10.7312/jenk19074-003}




\bibitem[Pe18]{penn2018ai}
Penn, J. AI thinks like a corporation | and that's worrying. \emph{The Economist}, November 26, 2018. \url{https://www.economist.com/open-future/2018/11/26/ai-thinks-like-a-corporation-and-thats-worrying}

\bibitem[Per22]{perrigo2022inside}
Perrigo, B. Inside Facebook's African Sweatshop. \emph{Time}, February 14, 2022. \url{https://time.com/6147458/facebook-africa-content-moderation-employee-treatment/}

\bibitem[Per23]{perrigo2023exclusive}
Perrigo, B. Exclusive: OpenAI Used Kenyan Workers on Less Than \$2 Per Hour to Make ChatGPT Less Toxic. \emph{Time}, January 18, 2023. \url{https://time.com/6247678/openai-chatgpt-kenya-workers/}
 
 
\bibitem[PF22]{pratt2022marginalized}
Pratt, A., Fowler, F. \emph{Deconstructing Bias: Marginalization}. June 2022, \url{https://science.nichd.nih.gov/confluence/pages/viewpage.action?pageId=242975243}


\bibitem[RHM22]{ray2022racializedburden}
Ray, V., Herd, P., Moynihan, D., Racialized Burdens: Applying Racialized OrganizationTheory to the Administrative State. \emph{Journal of Public Administration Research and Theory}, 2023, 33, 139-152, \url{https://doi.org/10.1093/jopart/muac001}

\bibitem[Re21]{regalado2021latinx}
Regalado, P. ``9. They Speak Our Language . . . Business": Latinx Businesspeople and the Pursuit of Wealth in New York City. In \emph{Histories of Racial Capitalism}, edited by Justin Leroy and Destin Jenkins, 231-250. New York Chichester, West Sussex: Columbia University Press, 2021. DOI: \url{https://doi.org/10.7312/jenk19074-011}

\bibitem[Ro83]{robinson1983black}
\label{robinson1983black}
Robinson, C. J. \emph{Black Marxism: The Making of the Black Radical Tradition}. University of North Carolina Press, 1983.

\bibitem[S14]{sassen2014expulsions}
Sassen, S. \emph{Expulsions: Brutality and Complexity in the Global Economy}. Harvard University Press, 2014, \url{https://doi.org/10.2307/j.ctt6wpqz2}.

\bibitem[WMG22]{williams2022exploited}
Williams, A., Miceli, M., \& Gebru, T. The Exploited Labor Behind Artificial Intelligence. \emph{Noema}, October 13, 2022. \url{https://www.noemamag.com/the-exploited-labor-behind-artificial-intelligence/}

\bibitem{young1990marginalized}
Young, I. Justice and the Politics of Difference, REV-Revised. Princeton University Press, 1990. \url{https://doi.org/10.2307/j.ctvcm4g4q.}




\end{thebibliography}
\end{document}